\documentclass[aps,,reprint,twocolumn,superscriptaddress,showpacs]{revtex4-1}
\usepackage{graphicx}
\usepackage{mathrsfs}
\usepackage{bm}
\usepackage{amsmath}
\usepackage{dcolumn}
\usepackage{epstopdf}
\usepackage{dsfont}
\usepackage{amssymb}
\usepackage{tabularx}
\usepackage{array}
\usepackage{float}
\usepackage{color}
\usepackage{epstopdf}
\usepackage{mathrsfs}
\usepackage[colorlinks, linkcolor=blue,anchorcolor=blue,citecolor=blue,urlcolor=blue]{hyperref}

\begin{document}

\title{Multiple magnetism controlled topological states in EuAgAs}

\author{Yahui Jin}\email{These authors contributed equally to this work.}
\affiliation{Key Laboratory of Low-Dimensional Quantum Structures and Quantum Control of Ministry of Education, Department of Physics and Synergetic Innovation Center for Quantum Effects and Applications, Hunan Normal University, Changsha 410081, China}

\author{Xu-Tao Zeng}\email{These authors contributed equally to this work.}
\affiliation{School of Physics, and Key Laboratory of Micro-nano Measurement-Manipulation and Physics, Beihang University, Beijing 100191, China}

\author{Xiaolong Feng}
\affiliation{Research Laboratory for Quantum Materials, Singapore University of Technology and Design, Singapore 487372, Singapore}

\author{Xin Du}
\affiliation{Key Laboratory of Low-Dimensional Quantum Structures and Quantum Control of Ministry of Education, Department of Physics and Synergetic Innovation Center for Quantum Effects and Applications, Hunan Normal University, Changsha 410081, China}

\author{Weikang Wu}
\affiliation{Research Laboratory for Quantum Materials, Singapore University of Technology and Design, Singapore 487372, Singapore}

\author{Xian-Lei Sheng}
\affiliation{School of Physics, and Key Laboratory of Micro-nano Measurement-Manipulation and Physics, Beihang University, Beijing 100191, China}

\author{Zhi-Ming Yu}
\affiliation{Key Lab of Advanced Optoelectronic Quantum Architecture and Measurement (MOE), Beijing Key Lab of Nanophotonics  and Ultrafine Optoelectronic Systems, and School of Physics, Beijing Institute of Technology, Beijing 100081, China}

\author{Ziming Zhu}
\email{zimingzhu@hunnu.edu.cn}
\affiliation{Key Laboratory of Low-Dimensional Quantum Structures and Quantum Control of Ministry of Education, Department of Physics and Synergetic Innovation Center for Quantum Effects and Applications, Hunan Normal University, Changsha 410081, China}

\author{Shengyuan A. Yang}
\affiliation{Research Laboratory for Quantum Materials, Singapore University of Technology and Design, Singapore 487372, Singapore}

\begin{abstract}
The interplay between magnetism and band topology is a focus of current research on magnetic topological systems. Based on first-principle calculations and symmetry analysis, we reveal that multiple intriguing topological states can be realized in a single system EuAgAs, controlled by the magnetic ordering. The material is a Dirac semimetal in the paramagnetic state, with a pair of accidental Dirac points. Under different magnetic configurations, the Dirac points can evolve into magnetic triply-degenerate points, magnetic linear and double Weyl points, or being gapped out and making the system a topological mirror semimetal characterized by mirror Chern numbers. The change in bulk topology is also manifested in the surface states, including the surface Fermi arcs and surface Dirac cones. In addition, the antiferromagnetic states also feature a nontrivial $\mathbb{Z}_4$ index, implying a higher order topology. These results deepen our understanding of magnetic topological states and provide new perspectives for spintronic applications.

\end{abstract}

\maketitle

\section{Introduction}
Topological states of matter have been attracting great interest in physics research \cite{hasan2010colloquium,qi2011topological,Chiu_RMP,Armitage_RMP}. Such states can be fully gapped in the bulk, as in topological insulators, where gapless excitations appear at boundaries in the form of protected boundary modes \cite{hasan2010colloquium,qi2011topological}. Meanwhile, in topological semimetals, unconventional quasiparticle excitations emerge in the bulk around protected band degeneracies \cite{Chiu_RMP,Armitage_RMP}. The degree of degeneracy directly determines the internal structure of the quasiparticles. For example, Weyl and Dirac points have twofold and fourfold degeneracy, respectively, and they give rise to quasiparticles analogous to Weyl and Dirac fermions in relativistic quantum field theory \cite{wan2011topological,burkov2011weyl,wang2012dirac,wang2013three,zhu2019composite}. The crystalline symmetry also allows degenerate points beyond the Weyl and Dirac types \cite{bradlyn2016beyond,yu2021encyclopedia}. The first example discovered is the triply degenerate point (also known as the triple point), which leads to three-component quasiparticles \cite{zhu2016triple,weng2016topological,chang2017nexus,lv2017observation,yang2019topological,xiaomingprb,zhong2017three}. Some of the topological semimetals also possess protected boundary modes, such as the surface Fermi arcs for Weyl semimetals \cite{wan2011topological,weng2015weyl,huang2015weyl}. These topological states have been extensively studied in nonmagnetic materials which respect the time reversal symmetry \cite{vergniory2019complete,tang2019comprehensive,zhang2019catalogue}.

Recently, the research has been expanded to magnetic systems \cite{xu2020high}. Due to the broken time reversal symmetry, some of the topological classifications are fundamentally changed and lead to rich topological states. Various topological insulating states, Weyl points, and Dirac points have been reported in magnetic materials \cite{xu2011chern,wan2011topological,hua2018dirac,zhang2019topological,li2019intrinsic,xu2019higher,gui2019new,jiao2019room,PhysRevB.100.064408,nie2020magnetic,xiaomingloop,zhu2020first}. Particularly, an important feature of magnetic topological systems is that the topology is sensitive to the magnetic configuration and may be conveniently controlled by tuning the magnetic order. For example, it was predicted that in magnetic materials such as EuCd$_2$As$_2$ \cite{hua2018dirac} and EuB$_6$ \cite{nie2020magnetic}, multiple topological states can be realized by rotating the magnetization direction or by switching the ordering between ferromagnetism and antiferromagnetism. The techniques for manipulating magnetism have been well developed in spintronics \cite{RevModPhys.76.323,hirohata2020review}, such that the close interplay between magnetism and band topology may generate interesting effects and open new perspectives for applications.

Regarding materials, compounds containing the lanthanide element Eu are appealing candidates, as Eu ions typically have large magnetic moments. The most desired case is that the magnetism is from the Eu $f$ orbitals, while the low energy bands around the Fermi level are from the dispersive $s$ or $p$ orbitals of other constituents, so that the topological features can be more clearly exposed and the carrier mobility can be enhanced.

In this work, we investigate the topological states in such a candidate material EuAgAs. EuAgAs single crystal was synthesized in 1981 \cite{tomuschat1981abx}. The magnetic property of EuAgAs has been studied in experiment and the material was shown to be metamagnetic, i.e., its magnetic configuration can be readily tuned by external means such as applied magnetic field \cite{tomuschat1984magnetische}. Here, based on first-principles calculations and symmetry/topology analysis, we show that
rich magnetic topological states can be realized in EuAgAs depending on the magnetic configurations. The high-temperature paramagnetic phase of EuAgAs is a Dirac semimetal with a pair of Dirac points. The ground state of EuAgAs is found to be antiferromagnetic (AFM), which is consistent with experimental result~\cite{tomuschat1984magnetische}. We show that the band topology is sensitive to the N\'{e}el vector direction. For N\'{e}el vector along the $c$-axis, the state has two pairs of triply degenerate points (TDPs), which were not observed in magnetic materials before. When the N\'{e}el vector lies in the $ab$-plane, the system is a special topological mirror semimetal that is adiabatically connected to a topological crystalline insulator characterized by two mirror Chern numbers and Dirac type surface states. In addition, we find that the ferromagnetic (FM) state of EuAgAs, which may be achieved by strain or applied magnetic field, possesses four pairs of Weyl points. Interestingly, one pair belongs to the double Weyl points, which feature quadratic band splitting in a plane and Chern numbers of $\pm 2$. The new states reported here, including the
magnetic TDPs, the magnetic topological mirror semimetals, and the magnetic double Weyl points, enrich the family of topological states in magnetic systems. Our work also suggests EuAgAs as a promising platform to explore the interplay between magnetism and band topology.

\section{Calculation methods}

Our first-principles calculations were based on  the density functional theory (DFT), using a plane-wave basis set and projector augmented wave method ~\cite{PhysRevB.50.17953}, as implemented in the Vienna \emph{ab} \emph{initio} simulation package (VASP) ~\cite{PhysRevB.54.11169,kresse1999ultrasoft}. The generalized gradient approximation (GGA) parameterized by Perdew, Burke, and Ernzerhof (PBE) was adopted for the exchange-correlation functional~\cite{PhysRevLett.77.3865}. We have also tested other functionals including PBEsol \cite{PhysRevB.79.155107} and SCAN ~\cite{PhysRevLett.115.036402,sun2016accurate}, which gave qualitatively the same results.
The energy cutoff was set to 360 eV, and a $11\times11\times5$ Monkhorst-Pack $k$ mesh was used for the Brillouin zone (BZ) sampling.  The atomic positions were fully optimized until the residual forces were less than $10^{-3}$ eV/{\AA}. The convergence criterion for the total energy was set to be $10^{-6}$ eV. Spin-orbit coupling (SOC) was included for all the results presented in the paper. In simulating the high-temperature paramagnetic state, the usual ``open core'' treatment of 4\emph{f} electrons was adopted. To account for the correlation effects of the \emph{f}-electrons in the magnetic phases, we have adopted the GGA+Hubbard-U (GGA+U) method \cite{PhysRevB.57.1505} with the value of $U = 8$ eV, which was commonly used in studying Eu compounds. We have also tested the $U$ values from 3 to 8 eV and found qualitatively consistent results.  The surface states were investigated by using the iterative Green's function method ~\cite{sancho1985highly} as implemented in the WannierTools package~\cite{wu2017wanniertools}.

\section{Crystal structure and magnetism}

\begin{figure}[!htp]
	{\includegraphics[clip,width=8.2cm]{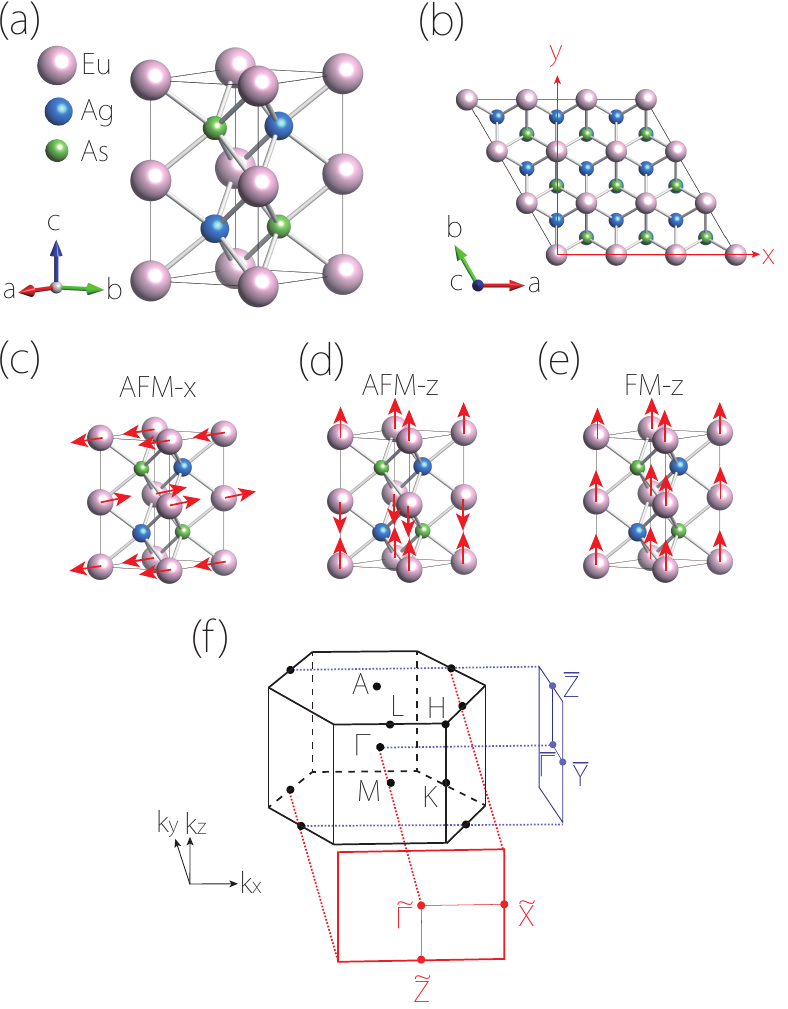}}
	\caption{\label{fig1}
		 (a) Unit cell of EuAgAs. (b) Top view of the EuAgAs lattice, where the $x$ and $y$ axis are defined.
(c-e) show the three magnetic configurations discussed in the main text. Red arrows indicate the local moment direction. (f) Bulk and surface BZ for EuAgAs.}
\end{figure}

EuAgAs crystallizes in the modified Ni$_2$In-type hexagonal structure with space group $P6_3/mmc$ (No. 194) as shown in Fig.~\ref{fig1}. In the structure, Ag and As together form planar honeycomb layers parallel to the $ab$ plane. In each honeycomb layer, the A and B sites are occupied by different atoms. These layers are stacked along the \emph{c} direction, with an alternation in the type of atoms at A and B sites.  Eu atoms are intercalated between the adjacent Ag-As layers. In a unit cell, Eu is at the Wyckoff position of 2\emph{a}, and Ag (As) is at the position of 2\emph{d} (2\emph{c}).
The experimental lattice constants with $a=4.516$ {\AA}  and $c=8.107$ {\AA} were adopted in our calculations ~\cite{tomuschat1981abx}.

The lattice symmetry group contains the following generators: threefold rotation $C_{3z}$, inversion $\mathcal{P}$, twofold rotation $C_{2y}$, and twofold screw rotation  $S_{2z}=\{C_{2z}|00\frac{1}{2}\}$ with a half translation along the $c$ ($z$) direction [the directions are labeled in Fig.~\ref{fig1}(b)]. Combining inversion with the twofold rotations, we also have three mirror symmetries $M_z$, $M_x$, and $\tilde{M}_y$, where
$\tilde{M}_y=\{M_y|00\frac{1}{2}\}$ is a glide mirror.

Magnetic properties of EuAgAs have been experimentally studied in Ref. \cite{tomuschat1984magnetische}. It was shown that the ground state is AFM with a N\'{e}el temperature about 11 K. The measurement showed that the magnetic moments are mainly on the Eu sites with a magnitude of $\sim 7.45$ $\mu_B$, which is consistent with the Eu$^{2+}$ ions being in the high-spin state of 4f$^{7}$. From experiment \cite{tomuschat1984magnetische}, it was found that the AFM is of A-type, i.e., the coupling within each Eu hexagonal layer is FM,  whereas the coupling between two neighboring layers along $c$ axis is AFM (see Fig.~\ref{fig1}(c) and (d)). Nevertheless, the measurement in Ref. \cite{tomuschat1984magnetische} cannot resolve the direction of the N\'{e}el vector. In Table ~\ref{table:msg}, we compare the calculated energies of the different magnetic configurations. It shows that consistent with experiment, the AFM state is energetically favorable. Among the AFM configurations, the N\'{e}el vector prefers to stay in the $ab$ plane, with small in-plane anisotropy. The AFM-$x$ state with  N\'{e}el vector along $x$ has the lowest energy, while the energies for the AFM-$y$ and AFM-$z$ states are also quite close. Experiments found that EuAgAs is a metamagnetic material \cite{tomuschat1984magnetische}, indicating that the magnetic configuration is sensitive to external perturbations, particularly the applied magnetic field. Hence, the FM configuration is also considered here. The different magnetic configurations may be switched by applied strain, magnetic field, or electric current. In the following sections, we shall analyze the topological features of each magnetic configuration in Table~\ref{table:msg}. Since the topological features in Table~\ref{table:msg} all occur on the $\Gamma$-$A$ path in the BZ. In Table~\ref{table:CT}, we give the character table of little groups on this path for different magnetic configurations, which will be useful for later discussions.

\begin{table}
	\centering
	\caption{ {Comparison of different magnetic configurations for EuAgAs. Here, MSG stands for the magnetic space group, we show the energy comparison obtained by three exchange-correlation functionals, the energy has unit of meV per formula unit and is with reference to the AFM-$x$ configuration, and the last column indicates the topological feature of each state.
DP, TMS, and WP stand for the Dirac point, topological mirror semimetal, and Weyl point, respectively. All the AFM configurations also carry a nontrivial $\mathbb{Z}_4$ index $\kappa=2$. }}
	\label{my-label}
	\renewcommand\arraystretch{1.4}
	 \begin{tabular}{p{0.0cm}<{\centering}p{1.2cm}<{\centering}p{1.5cm}<{\centering}p{1.0cm}<{\centering}p{1.0cm}<{\centering}p{1.5cm}<{\centering}p{1.5cm}<{\centering}p{1.5cm}<{\centering}p{1.5cm}<{\centering}}
		\hline\hline
		\rule{0pt}{16pt}
		&~~       &MSG    & PBE  & PBEsol   & SCAN  & Topology  \\
		\hline
		&PM &$P6_3/mmc$  &  & &   & DP \\
				
		& AFM-$x$ &$Cmcm$  &$0$ &$0$ &$0$  & TMS     \\
		
		& AFM-$y$ &$Cm'c'm$  &$0.014$ &$0.017$  &$0.038$   & TMS  \\
		
		& AFM-$z$ &$P6_3'/m'm'c$  &$0.163$ &$0.098$ &$0.153$   & TDP   \\
		
		& FM-$z$ &$P6_3/mm'c'$  &$6.788$ &$12.732$ &$8.975$ & WP   \\
		\hline\hline
	\end{tabular}
	\renewcommand\arraystretch{1.4}
	\label{table:msg}
\end{table}

\begin{table}
	\centering
	\caption{{Character table of little group on the $\Gamma$-$A$ path for the PM ($C_{6v}$), AFM-$x$ ($C_{2v}$), AFM-$y$ ($C_2$), AFM-$z$ ($C_{3v}$) and FM-$z$ ($C_6$) states. Here, $\omega=\exp(i\pi/6)$. }}
	\renewcommand\arraystretch{1.4}
	\begin{tabular}{p{0.0cm}<{\centering}p{2.0cm}<{\centering}p{0.9cm}<{\centering}p{0.9cm}<{\centering}p{0.9cm}<{\centering}p{0.9cm}<{\centering}p{0.9cm}<{\centering}p{0.9cm}<{\centering}}
		\hline\hline
		\rule{0pt}{16pt}
		&~Little Group&   &$E$    &$C_2$ &$C_3$ &$C_6$    &$\sigma_v$  \\
		\hline
		&$C_{6v}$	&$~\Gamma_{7}$    &2   &0 &1     &$\sqrt{3}$ &0          \\
		&&$~\Gamma_{9}$   &$2$   &0 &-2      &0    &0\\
		\hline
		&$C_{2v}$  &$~\Gamma_5$ &2 &0 &&&0 \\	
		\hline
		&$C_2$   	&$~\Gamma_3$ &1 &$i$ &    \\
		&	&$~\Gamma_4$ &1 &-$i$ &  \\
		\hline
		&$C_{3v}$     	&$~\Gamma_4$ &2 & &1&&0    \\
		&	&$~\Gamma_5$ &1 &&-1& &$i$    \\
		&	&$~\Gamma_6$ &1 &&-1& &-$i$    \\
		\hline
		&$C_6$ 	&$~\Gamma_7$ &1 &$i$&$\omega^2$ &$\omega$ &     \\
		&	&~$\Gamma_8$ &1 &-$i$&$-\omega^4$  &-$\omega^5$  & \\
		&	&$\Gamma_{11}$ &1 &$i$&-$1$& -$i$ &    \\
		&	&$\Gamma_{12}$ &1&-$i$ &-$1$ &$i$ &   \\
		\hline\hline
	\end{tabular}
	\renewcommand\arraystretch{1.4}
	\label{table:CT}
\end{table}

\begin{figure}[!tb]
	{\includegraphics[clip,width=8.2cm]{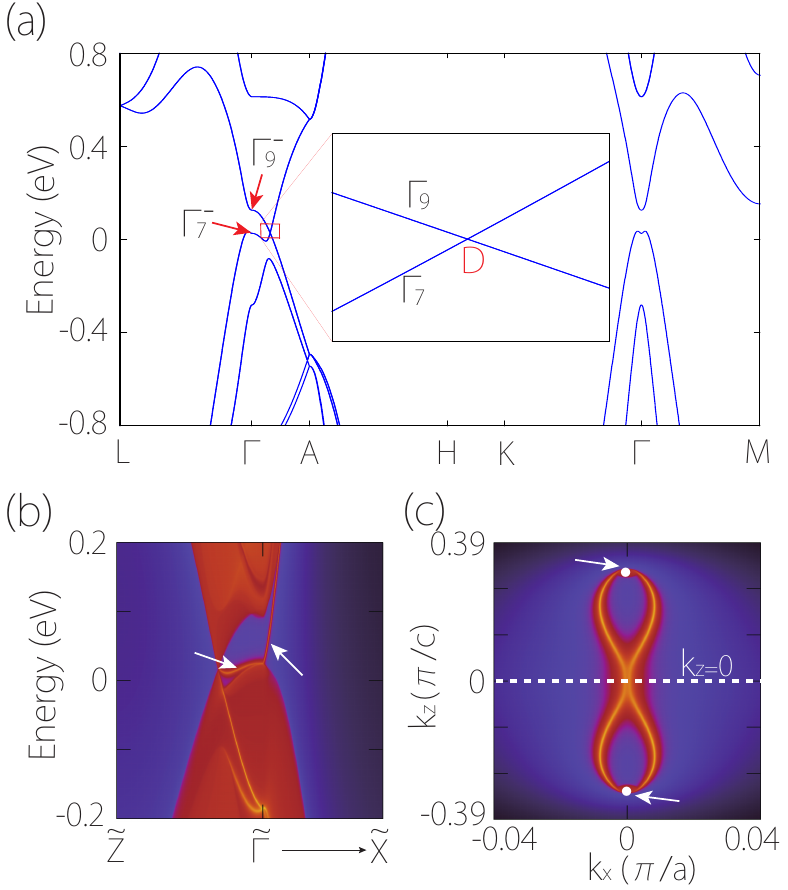}}
	\caption{\label{fig2}
		(a) Band structure of EuAgAs in the paramagnetic phase. The inset shows the enlarged view of the red box. The point $D$ is a fourfold Dirac point. (b) Calculated surface spectrum for the surface normal to $y$. Arrows indicate the surface states. (c) Constant energy slice of the surface spectrum at the energy of bulk Dirac points. The white dots indicate the surface projections of Dirac points.}
\end{figure}

\section{Paramagnetic state: Dirac semimetal}

Let's first consider the paramagnetic phase of EuAgAs, which occurs when the temperature is above the magnetic transition temperature.
The calculated band structure (with SOC included) is plotted in Fig.~\ref{fig2}. The system shows a semimetal character.
Interestingly, one observes that close to the Fermi level, the conduction and the valence bands cross each other at a point $D$ with coordinate $(0,0,k_D)$ ($k_D$= 0.316 in unit of  $\pi/{c}$) on the $\Gamma$-$A$ path. Here, since the system preserves both the time reversal and the inversion symmetries, each band is doubly degenerate. It follows that the crossing point $D$ is a fourfold degenerate Dirac point. From symmetry point of view, the little co-group on the $\Gamma$-$A$ path is $C_{6v}$, and the two crossing bands belong to two different two-dimensional irreducible representations $\Gamma_9$ and $\Gamma_7$ of $C_{6v}$, as indicated in Fig.~\ref{fig2}(a).  This shows that the Dirac point is symmetry protected.

Based on the symmetry, we derive the following $k\cdot p$ effective model for the states around the Dirac point $D$ in Fig.~\ref{fig2}(a):
\begin{align}
H_{D}({\bm q}) = a_0 q_z+a_1 (\sigma_y \tau_z q_x - \sigma_x q_y) + a_2 \sigma_z q_z,
\end{align}
where the energy and the momentum $\bm q$ are measured from the Dirac point, $\sigma_i$ and $\tau_i$ are Pauli matrices. The model parameters $a_i$ can be extracted from fitting the DFT band structure. The obtained values are $a_0$= 0.49 eV{\AA},
$a_1$  = 3.25 eV{\AA}, $a_2$= 2.10 eV{\AA}.
This model clearly demonstrates the character of a Dirac point. There is another Dirac point $D'$ at $(0,0,-k_D)$, related to $D$ by $\mathcal{P}$ or $\mathcal{T}$ symmetry.

This pair of Dirac points belong to the accidental band degeneracies. Their existence requires the local band inversion at $\Gamma$ and can be removed without breaking the symmetry of the system, e.g., by switching the order of  $\Gamma_{7}^{-}$ and $\Gamma_{9}^{-}$ states at $\Gamma$ [see Fig.~\ref{fig2}(a)]. Due to the band inversion, the 2D slice $k_z=0$ features a nontrivial $\mathbb{Z}_2$ invariant $\nu=1$, which is verified by Wilson loop method and parity analysis \cite{fu2007topological_p,PhysRevB.89.155114}. It follows that there must exist a time reversal pair of surface states on the $k_z=0$ line in the surface BZs for side surfaces. In Fig.~\ref{fig2}(b) and (c), we show the surface spectrum for the surface normal to $y$. One pair of surface Fermi arcs connecting the surface projections of Dirac points are spotted, similar to the cases of Na$_3$Bi \cite{wang2012dirac} and Cd$_3$As$_2$ \cite{wang2013three}.

\section{AFM-z state: magnetic TDP}

\begin{figure}[!tb]
	{\includegraphics[clip,width=8.2cm]{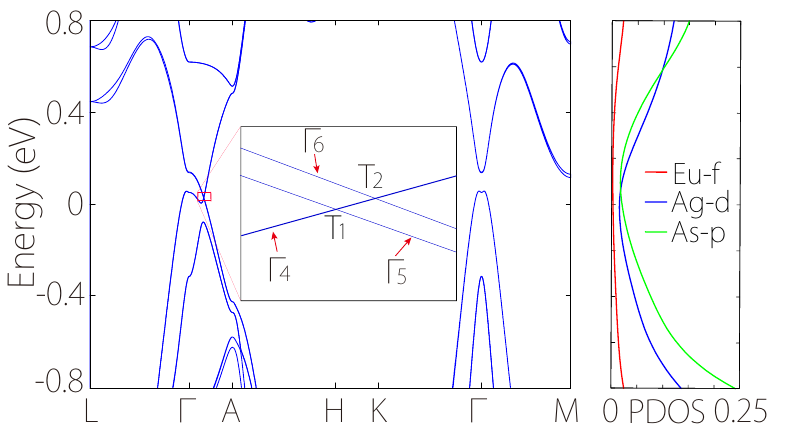}}
	\caption{\label{fig3}
		 Band structure and projected density of state (PDOS) for EuAgAs in the AFM-$z$ state. The inset shows an enlarged view of the red box. $T_1$ and $T_2$ are TDPs. }
\end{figure}

Next, we consider EuAgAs in the AFM-$z$ state. The calculated band structure is plotted in Fig.~\ref{fig3}. The overall structure is similar to the paramagnetic state. However, since the magnetic ordering breaks the time reversal symmetry, the original Dirac points will no longer be stable. Interestingly, from the inset of Fig.~\ref{fig3}, one observes that the Dirac point $D$ splits into two TDPs $T_1$ and $T_2$ on the $\Gamma$-$A$ path. Each TDP is a crossing between a doubly degenerate band and a non-degenerate band. Here, the doubly degenerate band belong to the $\Gamma_4$ representation of the $C_{3v}$ group, while the non-degenerate band corresponds to the $\Gamma_5$ ($\Gamma_6$) representation for $T_1$ ($T_2$).
TDPs have been extensively studied in nonmagnetic systems before \cite{zhu2016triple,weng2016topological,chang2017nexus,lv2017observation,yang2019topological,gao2018possible}, and their existence in materials MoP and WC have been experimentally verified \cite{lv2017observation,yang2019topological}. In comparison, the TDPs here are realized in a magnetic system, and we find that they can be derived from Dirac points via breaking the time reversal symmetry.
Based on the symmetries $C_{3v}$ on the $\Gamma$-$A$ path, we derive the following $k\cdot p$ effective model for $T_1$:

\begin{align}\nonumber
H_{T_{1}}({\bm q}) = & b_0 \mathbb{I}_3 q_z+(b_1 \Lambda_3 + b_2 \Lambda_6 + b_3 \Lambda_7)q_x \\ \nonumber
+ & (b_1 \Lambda_1 + b_4 \Lambda_4 + b_5 \Lambda_5)q_y \\
+ & b_6 \Lambda_8 q_z,
\end{align}
where the energy and the momentum are measured from $T_1$, $\Lambda_{i}$'s are the $3\times3$ Gell-Mann matrices (see Appendix ~\ref{gell} for their concrete forms), $\mathbb{I}_3$ is the $3\times 3$ identity matrix, and $b_i$'s are real parameters.
The obtained model for $T_2$ is quite similar, taking the form of
\begin{align}\nonumber
H_{T_{2}}({\bm q}) = & c_0 \mathbb{I}_3 q_z+(c_1 \Lambda_3 + c_2 \Lambda_4 + c_3 \Lambda_5)q_x \\ \nonumber
+ & (c_1 \Lambda_1 + c_4 \Lambda_6 + c_5 \Lambda_7)q_y \\
+ &  c_6 \Lambda_8 q_z,
\end{align}
where the energy and the momentum are measured from $T_2$.

We note that some recent works reported non-Abelian topology in three-band spinless systems~\cite{wu2019non, PhysRevB.103.L121101}. In comparison, the triple points here (also a three-band system) exist only when SOC is considered. Whether similar non-Abelian physics can emerge in spinful systems is an interesting question to explore in future studies.

\section{AFM-x state: topological mirror semimetal}

\begin{figure}[!tb]
	{\includegraphics[clip,width=8.2cm]{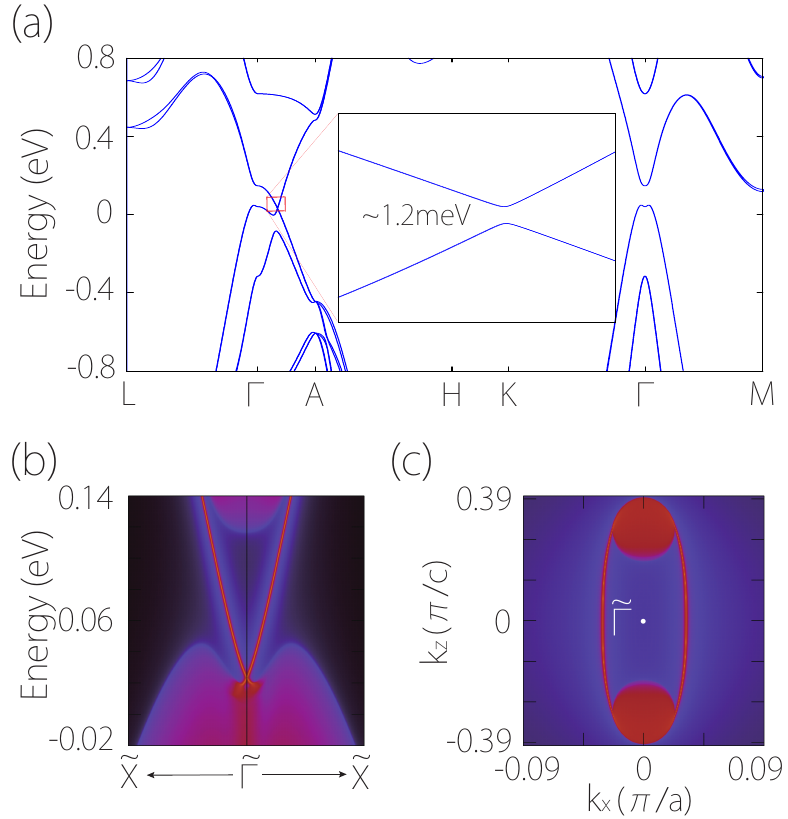}}
	\caption{\label{fig4}
		(a) Band structure of EuAgAs in the AFM-$x$ state. The inset shows an enlarged view of the red box. A local gap exists between conduction and valence bands throughout the BZ.  (b) Spectrum for the surface normal to $y$. (c) Constant energy slice for the surface spectrum at 60 meV. }
\end{figure}

When the N\'{e}el vector is oriented along the $x$ direction, besides the time reversal symmetry, the rotational symmetry along $z$ will also be broken. It turns out that all the crossings between conduction and valence bands will be destroyed, and the two bands are detached from each other. As shown in Fig.~\ref{fig4}(a), the original Dirac point in the paramagnetic phase is indeed removed, and a local band gap is maintained throughout the BZ, although there is a small indirect overlap in energy between the two bands. This means that the band structure is adiabatically connected to an insulator state. Note that the band inversion feature at $\Gamma$ is still maintained, so this semimetal state is in fact topological.

As the AFM-$x$ state preserves the mirror symmetries: $M_{x}$, $\tilde{M}_y$, and $M_z$, the band inversion at $\Gamma$ may be characterized by the mirror Chern numbers \cite{fu2011topological,hsieh2012topological,PhysRevB.78.045426} for the three mirror planes $k_{i}=0$ in the BZ, with $i\in\{x,y,z\}$. Here, the mirror Chern number is defined as
\begin{equation}
\mathcal{C}_{M_i}=(\mathcal{C}_{+}^{M_i}-\mathcal{C}_{-}^{M_i})/2,
\label{mcn}
\end{equation}
where $\mathcal{C}_\pm$ is the Chern number for valence bands with $\pm$ eigenvalue for the mirror $i$. From DFT calculations, {{we find that the mirror Chern number is indeed nontrivial  for the three mirror planes, capturing the band inversion at $\Gamma$.}}

Due to the nontrivial mirror Chern numbers, there will be Dirac type surface states for surfaces that respect at least one of the three mirrors. This is similar to the nonmagnetic topological crystalline insulator SnTe \cite{hsieh2012topological}. In Fig.~\ref{fig4}(b) and (c), we plot the surface spectrum for the surface normal to $y$.
We can see that Dirac type surface states appear on the surface, determined by the mirror Chern numbers. The Dirac cone is located at the $\Gamma$ point, as constrained by the presence of both $M_z$ and $M_x$ symmetries for this surface.

\begin{figure}[!tb]
	{\includegraphics[clip,width=8.2cm]{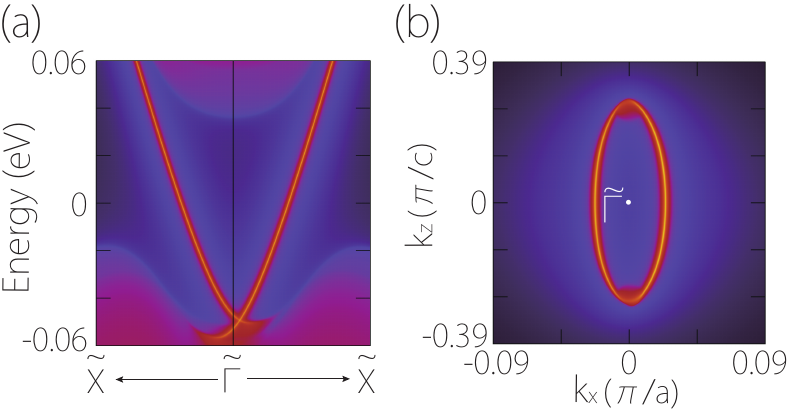}}
	\caption{\label{fig5}
		(a)  Spectrum for the surface normal to $y$ in the AFM-$y$ state. (b) Constant energy slice for the surface spectrum at 0 meV.}
\end{figure}

As for the AFM-$y$ state, the bulk band structure is very similar to the AFM-$x$ state in Fig.~\ref{fig4} (see Appendix ~\ref{AFM_y}). There is also a local gap maintained throughout the BZ, which separates the conduction and valence bands. In this case, both $M_x$ and $\tilde{M}_y$ are broken, but $M_z$ is still preserved and so the nontrivial $\mathcal{C}_{M_z}$ remains. Therefore, the AFM-$y$ state is also a topological mirror semimetal. Different from the AFM-$x$ state, here, the protected Dirac surface states only appear on side surfaces that preserve $M_z$. In Fig.~\ref{fig5}, we plot the spectrum for the surface normal to $y$. One can observe the surface Dirac cone. Notably, since there is only a single mirror $M_z$, the surface Dirac point deviates from the $\Gamma$ point and the cone is slightly tilted.

\section{FM-z state: magnetic double Weyl point}

\begin{figure}[!tb]
	{\includegraphics[clip,width=8.2cm]{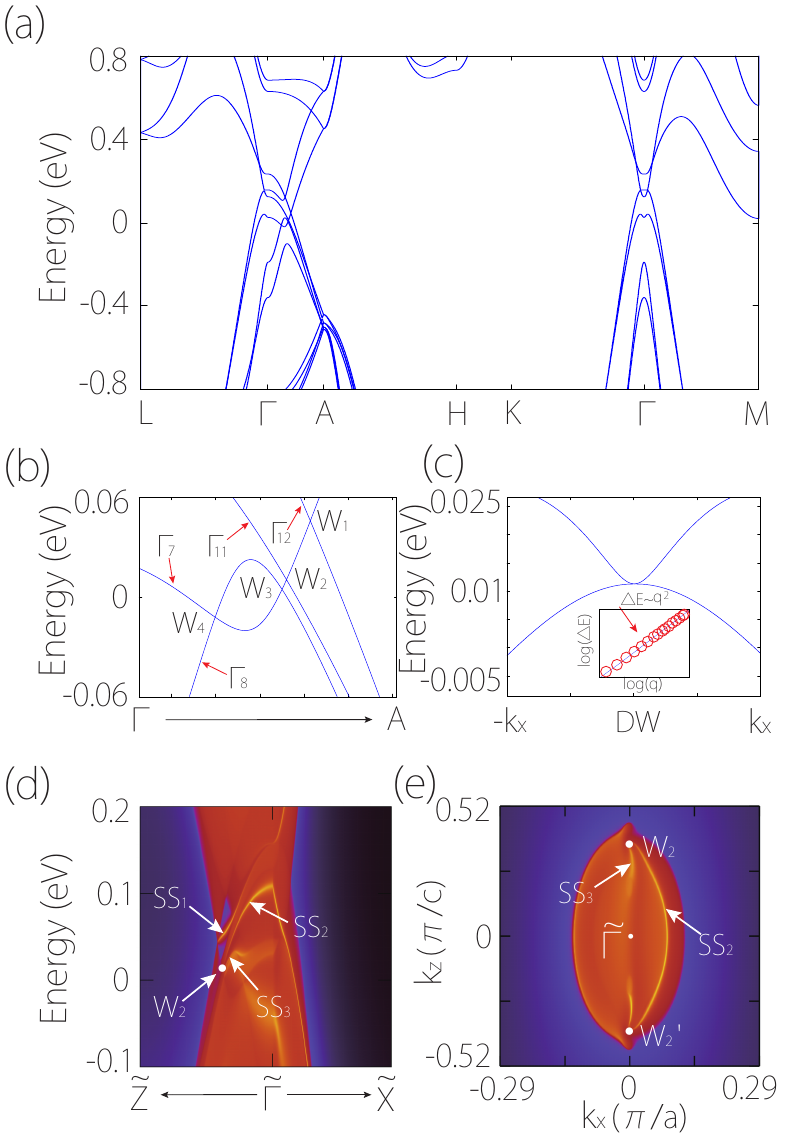}}
	\caption{\label{fig6}
		(a) Band structure of EuAgAs in the FM-$z$ state. (b) shows an enlarged view of the low-energy part on the $\Gamma-A$ path. Four Weyl points $W_i$ are labeled. $W_2$ is a double Weyl point. (c) shows the dispersion around the double Weyl point $W_{2}$ along the $k_x$ direction. The inset shows the log-log plot of the band splitting $\Delta E$ between the two bands versus $k_x$, indicating the quadratic character.
		(d) Surface spectrum for the surface normal to $y$. Location of the projected $W_2$ is labeled. (e) Constant energy slice taken at the energy of $W_2$. One can observe two Fermi arcs emanating from each of the projected double Weyl points. }
\end{figure}

Finally, we investigate the band structure for the FM-$z$ state. As shown in Fig.~\ref{fig6}(a), due to the exchange splitting, the band structure are quite different from the cases with paramagnetism or AFM. The state is more metallic, with enhanced density of states at Fermi energy. Interestingly, the low energy bands exhibit multiple crossings on the $\Gamma$-$A$ path, as indicated in Fig.~\ref{fig6}(a). Since the bands are non-degenerate, these crossing points are twofold Weyl points. We have checked that the points $W_{1}$, $W_{3}$ and $W_{4}$ are (linear) Weyl points with chirality of $-1$, $+1$ and $-1$, as shown in Table \ref{table:wc}. Remarkably, the point $W_{2}$ is found to be a double Weyl point with a topological charge of $+2$. The band splitting around $W_{2}$ is linear along $k_z$ but quadratic in the plane perpendicular to $k_z$. Such double Weyl points have been extensively studied in nonmagnetic systems \cite{xu2011chern,fang2012multi,huang2016new}. In magnetic systems, they are so far reported only in ferromagnetic HgCr$_2$Se$_4$ \cite{xu2011chern}.

In the current case, the double Weyl point is stabilized by the $C_{6}$ symmetry. We derive the following effective model for the states around $W_{2}$:
\begin{align}\nonumber
H_{DW}({\bm q}) = & d_0 q_z+d_1 \sigma_z q_z + (d_2 + d_3 \sigma_z)q_+ q_- \\
+ & \alpha q_{-}^2\sigma_{+} + \alpha^* q_{+}^2\sigma_{-},
\end{align}
where the $d_{i}$'s are real parameters, $\alpha$ is a complex parameter, $q_{\pm} = q_{x} \pm iq_{y}$, and $\sigma_{\pm} = \sigma_x \pm i\sigma_y$.

It is well known that the Weyl points will generate surface Fermi arcs that connect the their projections on a surface. In Fig.~\ref{fig6}(d) and (e), we plot the spectrum for the surface normal to $y$. Due to the overlap with the bulk bands, the pattern of the surface states is a bit difficult to distinguish. Nevertheless,  one can still observe that two Fermi arcs are emanating from each projected double Weyl point on the surface, consistent with the doubled topological charge.

The discussion above is for the FM-$z$ state. When the magnetization direction deviates from the $z$ direction, we expect that the linear Weyl points will still be present, as each Weyl point is topologically protected, but their locations will change. Meanwhile, the double Weyl points will no longer be stable, as they require the six-fold axis which is broken by the deviation. As a result, each double Weyl point will generally split into two linear Weyl points, during which the total topological charge should be conserved.

\begin{table}
	\centering
	\caption{{Weyl points in Fig.~\ref{fig6}(b) for EuAgAs in the FM-$z$ state. Here, the location $k_z$ is in unit of $\pi/c$ and energy is in unit of meV.}}
	\label{my-label}
	\renewcommand\arraystretch{1.4}
	 \begin{tabular}{p{0.0cm}<{\centering}p{1.75cm}<{\centering}p{1.25cm}<{\centering}p{1.25cm}<{\centering}p{2.25cm}<{\centering}p{2.25cm}<{\centering}p{1.4cm}<{\centering}}
		\hline\hline
		\rule{0pt}{16pt}
		& Weyl node         &$k_z$  &$E$   & Chern number  \\
		\hline
		&$~W_1$   &$0.4112$   &$45.9$     &$-1 $  \\
		
		&$~W_2$  &$0.3680$   &$10.6$      &$+2$  \\
		
		&$~W_3$    &$0.3604$   &$4.8$      &$+1$  \\
		
		&$~W_4$   &$0.2392$   &$-12.0$       &$-1 $  \\
		\hline\hline
	\end{tabular}
	\renewcommand\arraystretch{1.4}
	\label{table:wc}
\end{table}

\section{Discussion and conclusion}

In this work, using EuAgAs as an example, we demonstrate that a variety of interesting topological states can be realized in a single system, depending on the magnetic configurations. This also indicates that topological phase transitions can be induced by manipulating the magnetic ordering, for which the techniques have been under rapid development in field of spintronics. For example, the direction of N\'{e}el vector can be rotated by magnetic resonance, light pulse, or applied electric current~\cite{RevModPhys.76.323,hirohata2020review}. As EuAgAs is a metamagnetic material, its magnetic magnetism is sensitive to external perturbations. It is possible to change the magnetic ordering by applied strain or magnetic fields. For example, we find that by applying a uniaxial strain $\sim0.2$\% along $c$, the ground state can transform from  AFM-$x$ to AFM-$z$.

The various bulk and surface topological features predicted here can be directly imaged by the angle-resolved photoemission spectroscopy (ARPES). Scanning tunneling spectroscopy and magneto-transport may also be used to probe the band topology. We note that a very recent experiment~\cite{PhysRevB.103.L241112} reported a positive longitudinal magneto-conductivity in EuAgAs. As mentioned, under magnetic field, EuAgAs can be easily turned into the FM phase, which, according to our prediction, contains Weyl points near the Fermi level. The observed positive longitudinal magneto-conductivity is very likely to be attributed to these Weyl points, hence the result supports our theoretical prediction.

It is interesting to note that the inversion symmetry $\mathcal{P}$ is preserved for all the configurations in Table ~\ref{table:msg}. It follows that the nontrivial topology of the band structures in Figs. \ref{fig3}-\ref{fig5}  may also be captured by the parity analysis. Particularly, we may consider the $\mathbb{Z}_4$ index determined by the parity eigenvalues \cite{highorderTI2012}:
\begin{equation}
\kappa=\sum_{\Gamma_i}^{ } \frac{n^{\Gamma_i}_{+} -n^{\Gamma_i}_{-}}{2}\quad \bmod 4
 \label{z4}
\end{equation}
where $n^{\Gamma_i}_{\pm}$ denotes the number of valence bands with positive/negative parity at the inversion-invariant momentum point $\Gamma_i$.
From DFT calculations, we find that the $n^{\Gamma_i}_{\pm}$ values are unchanged for the paramagnetic and the AFM states, which can be readily understood since they share the same local band inversion at $\Gamma$. We find that all the AFM states here share the nontrivial $\kappa=2$. It has been shown that for insulators, $\kappa=2$ indicates a second order topology \cite{highorderTI2012}. As the AFM states with in-plane N\'{e}el vectors, e.g., the AFM-$x$ and AFM-$y$ states, are adiabatically connected to insulators, they are simultaneously a second-order topological semimetal, with topological hinge modes. For the paramagnetic state and the AFM-$z$ state, hinge modes are also expected to exist, similar to the recent study by Wieder et al. \cite{wieder2020strong}. However, as these states are semimetals, we find that the hinge modes overlap with the bulk bands and are difficult to distinguish in the calculated spectra.

Finally, we should mention that the example material EuAgAs is still far from ideal for exhibiting the discussed topological physics. There are a few drawbacks. First, the valence band has a hump along the $\Gamma$-$L$ path, which overlaps with the Dirac points and the TDPs in energy. Second, the locations of the two TDPs are quite close, which may pose difficulty for resolving them in ARPES experiment. Third, for cases such as the FM state, the projected bulk bands overlap with the surface states, which would interfere the experimental detection. Nevertheless, the knowledge we gained from this material is general and offers guidance to search for more suitable topological magnetic materials in future.

In conclusion, we have revealed that multiple interesting magnetic topological states can be realized in a single system by tuning the magnetic order. With EuAgAs as an example, we show that its paramagnetic, antiferromagnetic, and ferromagnetic states can manifest distinct topologies. Particularly, starting from a pair of accidental Dirac points in the paramagnetic state, a variety of states, including the magnetic TDPs, topological mirror semimetals, magnetic linear and double Weyl points, can be derived depending on how the symmetry is broken by the magnetic orders. The change in bulk topology is also accompanied with the change in surface states. Manipulating magnetism is a central task in spintronics. The interplay between magnetism and band topology revealed in our work can offer a new perspective for the study of spintronics.

\appendix

\section{\label{gell} Gell-Mann Matrices }

Gell-Mann matrices are traceless Hermitian generators of the SU(3) Lie algebra. Here, they are defined to take the following forms
\begin{align}\nonumber
\Lambda_1=&\begin{bmatrix}
0 & 1 & 0 \\
1 & 0 & 0 \\
0 & 0 & 0
\end{bmatrix},~\Lambda_2=\begin{bmatrix}
0 & -i & 0 \\
i & 0 & 0 \\
0 & 0 & 0
\end{bmatrix},~\Lambda_3=\begin{bmatrix}
1 & 0 & 0 \\
0 & -1 & 0 \\
0 & 0 & 0
\end{bmatrix}, \\\nonumber
\Lambda_4=&\begin{bmatrix}
0 & 0 & 1 \\
0 & 0 & 0 \\
1 & 0 & 0
\end{bmatrix},~\Lambda_5=\begin{bmatrix}
0 & 0 & -i \\
0 & 0 & 0 \\
i & 0 & 0
\end{bmatrix},~\Lambda_6=\begin{bmatrix}
0 & 0 & 0 \\
0 & 0 & 1 \\
0 & 1 & 0
\end{bmatrix},~\\\nonumber
\Lambda_7=&\begin{bmatrix}
0 & 0 & 0 \\
0 & 0 & -i \\
0 & i & 0
\end{bmatrix},~\Lambda_8=\frac{1}{\sqrt{3}}\begin{bmatrix}
1 & 0 & 0 \\
0 & 1 & 0 \\
0 & 0 & -2
\end{bmatrix}.
\end{align}
These matrices, together with the $3\times 3$ identity matrix, form a complete basis for $3\times 3$ Hermitian matrices.

\section{\label{AFM_y} Band structures for AFM-$y$ state}
\begin{figure}[H]
	{\includegraphics[clip,width=8.2cm]{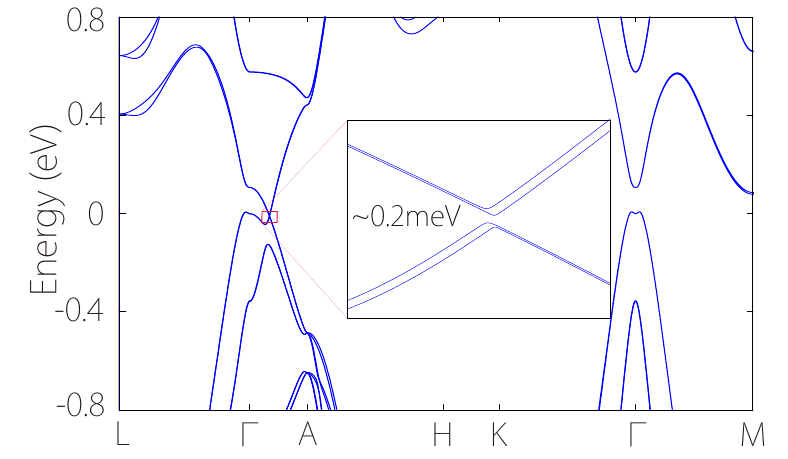}}
	\caption{\label{afm-y}
		Band structure of EuAgAs in the AFM-$y$ state. The inset shows the band crossing along $\Gamma$-A is fully gapped.}
\end{figure}

The calculated band structure for the AFM-$y$ state is shown in Fig.~\ref{afm-y}, which is similar with that of AFM-$x$ state (see Fig.~\ref{fig4} (a)).

\begin{acknowledgements}
The authors thank Quan-Sheng Wu, Mingda Li and D. L. Deng for valuable discussions. This work is supported by the National Natural Science Foundation of China (NSFC) (Grant No.~11704117 and 11974076) and the Singapore Ministry of Education Academic Research Fund Tier 2 (MOE2019-T2-1-001). We acknowledge computational support from Texas Advanced Computing Center, and H2 clusters in Xi'an Jiaotong University.
\end{acknowledgements}

\bibliography{MTS_refsv2}

\end{document}